\begin{filecontents*}{example.eps}
\RequirePackage{fix-cm}
\documentclass[smallextended]{svjour3}       % onecolumn (second format)
\smartqed  % flush right qed marks, e.g. at end of proof
\usepackage{graphicx}
\journalname{Experimental Astronomy}
\usepackage{array}
\begin{document}

\title{Performance of Large Area X-ray Proportional Counters in a Balloon Experiment}
\titlerunning{Performance of LAXPC in a Balloon Experiment}

\author{J. Roy$^{1,2}$, P. C. Agrawal$^{1,2}$, D. K. Dedhia$^{1}$, R. K. Manchanda$^{1,3}$, P. B. Shah$^{1}$, V. R. Chitnis$^{1}$, V. M. Gujar$^{1}$, J. V. Parmar$^{1}$, D. M.Pawar$^{1}$, V. B. Kurhade$^{1}$}

\offprints{jayashree.roy@gmail.com, jayashree.roy@cbs.ac.in}
\institute{
Tata Institute of Fundamental Research, Homi Bhabha Road, Mumbai-400005, India\\
\and	
Part of the work is carried out at present address: UM-DAE Center for Excellence in Basic Sciences, Santacruz (east), Mumbai-400098, India\\
\and
Present address: Dept of Physics, University of Mumbai, Kalina Campus. Mumbai - 400098
}

\authorrunning{J. Roy et al.} 

\date{Received: date / Accepted: date}

\maketitle

%\setstretch{2.0}
\begin{abstract}
ASTROSAT is India's first satellite fully devoted to astronomical observations covering a wide spectral band from optical to hard X-rays by a complement of 4 co-aligned instruments and a Scanning Sky X-ray Monitor. One of the instruments is Large Area X-ray Proportional Counter with 3 identical detectors.  In order to assess the performance of this instrument, a balloon experiment with two prototype Large Area X-ray Proportional Counters (LAXPC) was carried out on 2008 April 14. The design of these LAXPCs was similar to those on the ASTROSAT except that their field of view (FOV) was 3$^{\circ}$ $\times$ 3$^{\circ}$ versus FOV of 1$^{\circ}$ $\times$ 1$^{\circ}$ for the LAXPCs on the ASTROSAT. The LAXPCs are aimed at the timing and spectral studies of X-ray sources in 3-80 keV region. In the balloon experiment, the LAXPC, associated electronics and support systems were mounted on an oriented platform which could be pre-programmed to track any source in the sky. A brief description of the LAXPC design, laboratory tests, calibration and the detector characteristics is presented here. The details of the experiment and background counting rates of the 2 LAXPCs at the float altitude of about 41 km are presented in different energy bands. The bright black hole X-ray binary Cygnus X-1 (Cyg X-1) was observed in the experiment for $\sim$ 3 hours. Details of Cyg X-1 observations, count rates measured from it in different energy intervals and the intensity variations of Cyg X-1 detected during the observations are presented and briefly discussed.

\end{abstract}

\section{Introduction}
\label{introduction}
ASTROSAT is a multi-wavelength astronomy satellite which will carry four co aligned X-ray and UV instruments and a Scanning Sky Monitor (SSM), for wide spectral band studies of cosmic sources (\cite{Agrawal 2006a}, \cite{Agrawal 2006b}). Out of the 5 instruments aboard the ASTROSAT, four are sensitive in the X-ray band and one instrument namely Ultraviolet Imaging Telescope (UVIT) covers Visible (320-530 nm), Near UV (180-300 nm) and Far UV (130-180 nm) bands. The 4 X-ray instruments are : (1) The LAXPC instrument, with an effective area of $\ge$ 6000 cm$^{2}$ covers 3-80 keV spectral band for high time resolution studies and low spectral resolution studies of X-ray sources. (2) The Soft X-ray Telescope (SXT) uses conical foils based X-ray reflecting mirrors with an X-ray sensitive Charge Coupled Detector (CCD) at the focal plane to record low resolution ($\sim$ 3 arc minutes angular resolution) X-ray images and with moderate resolution X-ray spectra in the 0.5-8 keV spectral band. (3) The Cadmium - Zinc - Telluride Imager (CZTI) with geometrical area of 1000 cm$^{2}$ and energy resolution of about 10\% at 60 keV  uses an array of CZT detectors and has a coded aperture mask placed above the detector plane for recording intensity and spectra of sources with  angular resolution of $\sim$ 8 arc minute  in the 10-100 keV hard X-ray band. (4) A Scanning Sky Monitor (SSM) consisting of 3 position sensitive proportional counters equipped with one dimensional coded aperture masks is aimed at studies of time variability of new transients and known X-ray sources in 2-10 keV interval. All the X-ray instruments except SSM and the UVIT are co-aligned and mounted on the top deck of the satellite for simultaneous observations of cosmic sources over the broad spectral band. For a more detailed description of ASTROSAT refer to \cite{Agrawal 2006a}.
In the following section we present a detailed description of the design of the LAXPC instrument and its characteristics.

\section{Description of the LAXPC Instrument used in the Balloon Experiment}
\label{description LAXPC instrument}
The LAXPC instrument on the ASTROSAT consists of 3 identical detectors with associated independent high voltage (HV) supply, front-end electronics and signal processing electronics. Two prototype LAXPC units similar in design to those on the ASTROSAT mission except with FOV of 3$^{\circ}$ $\times$ 3$^{\circ}$ were used in the balloon experiment. We briefly describe the details of the LAXPC detector, H.V. supply, front-end electronics and signal processing electronics below.
\subsection{LAXPC Detector and Front-end Electronics}
LAXPC detector is a multi-anode, multi-layer proportional chamber that has 60 anode cells arranged in 5 layers in anode frames which serve as the X-ray detection volume. This is shown schematically in Figure \ref{fig1}(a) and  an actual photograph in \ref{fig1}(b). Each anode cell has a cross-section of 3.0 cm $\times$ 3.0 cm and a length of 100 cm. The anode consists of a gold coated stainless steel wire of 37 $\mu$ diameter, which is held under tension by soldering it at either end to the silver plated brass pin held in Teflon insulators mounted in the anode frames. Each anode cell is electrically isolated from the neighboring cells by a wall of 50 $\mu$ diameter beryllium-copper wires spaced 3 mm apart. Thus the X-ray detection volume is 15 cm deep, 36 cm wide and 100 cm long to achieve high detection efficiency for X-rays of up to 80 keV. The X-ray detecting anode cells are surrounded on three sides by veto layers. The veto layer has 46 veto cells, 11 on each long side and 24 below the bottom anode layer. The veto layers consist of veto cells of 1.5 cm $\times$ 1.5 cm cross-section and use 37 $\mu$ gold plated stainless steel wire as the anode of the veto cells.

\begin{figure}
\includegraphics[width=10cm]{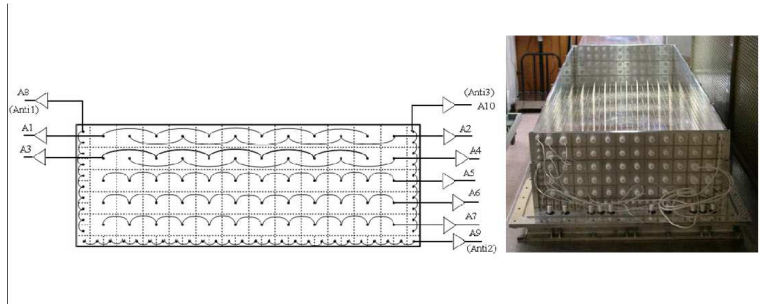}
\caption{1(a) Schematic diagram of anode cells of LAXPC detector arranged in 5 anode layers in anode frames which serve as the X-ray detection volume of the detector. 1(b) End-on view of the wired and assembled LAXPC anode frames.}
\label{fig1}      
\end{figure}
  
In the first 2 anode layers alternate anode cells are linked together to provide 2 outputs from each layer which are then operated in mutual anti-coincidence as shown in Figure \ref{fig1}(a). The anode cells in each of the other 3 layers are linked together to provide a single output from each of the 3 layers. Thus there are 7 outputs from the X-ray detector which are fed to seven independent Charge Sensitive Pre-amplifiers (CSPA's). All the 7 CSPA outputs are operated in mutual anticoincidence and an event is considered to be due to X-ray only if the pulse is detected from only one of the 7 outputs. However two simultaneous events in different anodes are accepted as X-ray events if the energy of one of the events falls in the energy band centered at 29.8 keV, being energy of Xenon K-alpha fluorescent X-ray. This is to ensure that genuine X-ray events of energy $>$ Xenon K-edge at 34.5 keV are not rejected as in 90\% of their interactions a fluorescent photon is emitted that may be absorbed in the detector volume in a different anode. The Veto cells on two sides and the bottom veto layer are linked together giving three veto layer outputs fed to three separate CSPA's. All the 3 veto layers are in anticoincidence with the 7 X-ray detection anode layers. This scheme ensures efficient rejection of non-X-ray background due to cosmic rays or events due to energetic electrons produced by Compton scattering of gamma-rays. Veto layers efficiently reject interactions of charged particles and gamma-rays in the walls of the anode frames and the detector housing. 
      For the purpose of data recording and transmission, the two outputs of each of the first two layers are combined after the logic circuitry so that there is one output from each layer. Similarly outputs of layers 3, 4 and 5 are combined to provide a single output from the 3 layers. This is aimed at simpler signal processing and data transmission scheme for the balloon experiment. Details of the combined outputs and their nomenclature are presented in Table \ref{wiringscheme.tab}.
 
\begin{table}
\begin{center}
\footnotesize
\caption{Wiring  and combination scheme of different layers of LAXPC detector}
\label{wiringscheme.tab}
\begin{tabular}{lll}
\hline         
Electronic output& Actual anode Layer & Anode Combination\\
\hline\\
Layer-A&Layer-1& A1 \& A2 (1A + 1B)\\
Layer-B&Layer-2& A3 \& A4 (2A + 2B)\\
Layer-C&Layer-3,4,5 Combined& A5, A6 \& A7 (3A + 4A + 5A) \\
	&	&Combined\\
ANTI-1&Left Anti& A8\\
ANTI-2&Bottom Anti& A9\\
ANTI-3&Right Anti& A10\\
ANTI&ANTI-1, ANTI-2, ANTI-3 Combined& A8 + A9 + A10 Combined\\[6pt]
\hline
\end{tabular}
\end{center}
\end{table}

The entire anode cell assembly is then mounted on a milled out Back Plate (BP) using fasteners tightened into blind-tapped holes. The BP has a milled out O-ring groove  on the top side to hold a O-ring as seen in Figure \ref{fig2}. The rear part of the BP is milled out to create pockets in which CSPA's, H.V. generation cards, distribution cards and command control electronics are housed. 
          The Detector Housing (DH) is placed over the Viton O-ring housed in the O-ring groove of the BP  to seal the detector on its bottom side. The DH is also milled out as a rectangular shaped enclosure open at two sides from a solid block of aluminium alloy. It has 2 mm thick walls which are stiffened by crossed ribs milled on the outside walls of the DH. This gives the DH required stiffness and strength to support 2 atmospheric pressure of the gas in the detector. Both the ends of the DH have milled out flanges with regularly spaced holes to mount the fasteners. The bottom flange sits in contact with the Viton O-ring while the top flange has a milled out O-ring groove for housing O-ring in it to seal the detector from the top side. A sandwich of 25 $\mu$ perforated polypropylene (PP) film and 50 $\mu$ thick Mylar sheet, aluminized on one side with 1000 \AA~thick aluminium, is placed in contact with the Viton O-ring to seal the detector on the top side. This is so placed that the PP film is in contact with the edges of the window support collimator so that the Mylar film is not damaged due to any rough edges in collimator slats. The Mylar sheet serves as a gas barrier as well as X-ray entrance window.

\begin{figure}
\includegraphics[width=10cm]{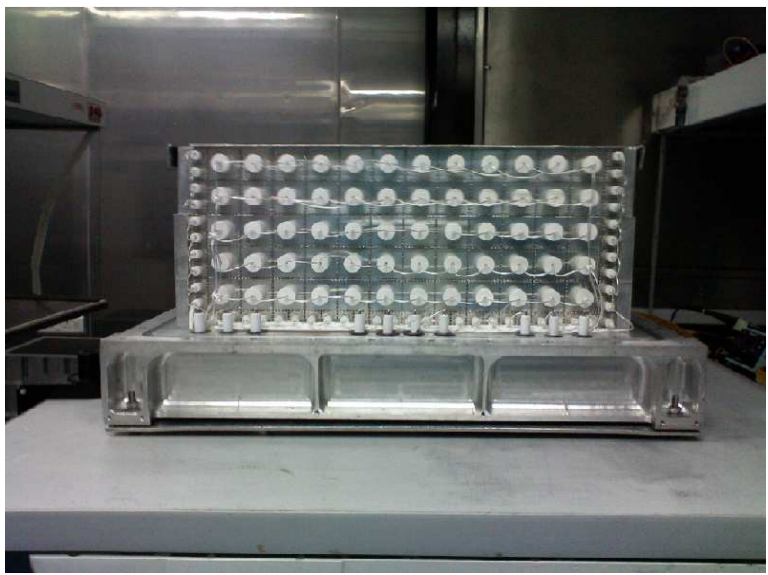}
\caption{LAXPC Anode Assembly mounted on the Back Plate}
\label{fig2}      
\end{figure}
There are two co-aligned collimators for each LAXPC that are mounted in a Collimator Housing (CH).
The CH structure is milled out from a single block of aluminium alloy with regularly spaced crossed ribs to provide stiffness and strength to the CH so that it can withstand one atmosphere of pressure difference during the evacuation and baking of the detectors. The CH has 3 milled out sections for mounting the collimators which are made in 3 identical parts. The collimators of LAXPC that support the thin Mylar film against the gas pressure as well as define the Field of View (FOV) of the LAXPC are mounted in the CH. The bottom flange of the CH has a milled out groove to mount a 1.5 mm cross-section Indium O-ring to make the CH leak proof. This is required during long bake out of the inside surfaces of the LAXPC when the CH is kept under vacuum prior to the filing of the detector with the counting gas. The CH is placed on top of the O-ring in the top flange of DH sandwiching the Mylar film between the Viton O-ring and the indium O-ring. The 6 mm diameter stainless steel fasteners are then placed in the bottom and top flanges and tightened to make the LAXPC assembly leak proof.

\subsection{The LAXPC Collimators}
The LAXPCs use 50 $\mu$ thick aluminized film as the X-ray window. The detector is filled with counting gas at about 2 atmosphere pressures. The PP and Mylar film sandwich rests against the smooth edges of a Window Support Collimator (WSC) that is housed in the lower part of the CH. The WSC is made of 3 identical sections mounted in a single CH. WSC has to withstand 2 atmosphere excess pressure in vacuum and hence it has to be made from high strength aluminium alloy. Each section of the WSC is made of 0.25 mm thick aluminium alloy slats of 7.5 cm width and appropriate length in which 0.40 mm wide slits are made along the width by laser cutting with center to center spacing of 7.0 mm. The slats are then assembled by crossing the slats to make square cell shaped sections of the collimator. This collimator has a field of view of 5$^{\circ}$ $\times$ 5$^{\circ}$. The Mylar film rests under pressure against the smooth edges of the slats of WSC. The WSC sits on the milled out lips milled in the inner walls of the CH and is glued to them by using a suitable epoxy.   
 The Field of View Collimator (FOVC) is also made in 3 identical sections. Each section is made from sheets of 5 layers of tin, copper and aluminium with total thickness of 175 $\mu$. The composite sheets are made by using 50 $\mu$ thick tin sheet, glued on either side with 12 $\mu$ thick copper which is then bonded to 50 $\mu$ thick aluminium sheets on either side of the copper surface. The bonded sheets of 15.5 cm width and appropriate lengths are then cut by a laser for making the collimator slats. Slits of 0.45 mm width are then cut by laser with center to center spacing of 7.0 mm. The slats are then baked at 70 $^{\circ}$C for several days in a vacuum chamber to completely cure and degas the epoxy used in the bonding of the composite sheet. The slats are finally crossed and assembled by gluing with a low out-gassing epoxy to make sections of the collimator assembly with square cell shaped geometry. The FOVC sits on the top side of the lips milled out along the inner walls of the CH. Thus FOVC is placed just above the WSC such that each section of WSC and FOVC is aligned. The two collimators are aligned carefully cell by cell so that cells of FOVC lie exactly above that of the WSC. Finally the two collimators are bonded to the inner walls and lips of the collimator housing using a suitable epoxy. The LAXPCs to be used in the ASTROSAT mission will have FOV of about 1$^{\circ}$ $\times$ 1$^{\circ}$. The LAXPC collimators used in the balloon experiment had FOV of 3$^{\circ}$ $\times$ 3$^{\circ}$  due to coarse pointing accuracy of about 0.5$^{\circ}$  of oriented gondola based on the use of a magnetic flux gate sensor.

\subsection{LAXPC High Voltage Unit}
A  new command controllable high voltage unit that can supply voltage in the range of 1.5 kV to about 3.5 kV and whose output can be varied by command in steps  of about 10 V or higher was used. A detailed description of the HV unit is presented elsewhere. The HV unit has a corona sensing part that shuts down the HV unit if there is indication of corona. The HV unit can be switched on and off by radio command. The HV unit also has provision for automatic lowering of HV to a nominal value when it encounters zones of high counting rates such as in the South Atlantic Anomaly (SAA) region in near earth inclined orbits.  There is also provision for lowering or restoration of HV by radio command.

\subsection{The Signal processing Electronics}
The signal processing electronics is designed to accept all the events that are (i) within 3-80 keV band (ii) not accompanied by a veto layer pulse (iii) satisfy mutual anticoincidence condition i.e. only one pulse at a given time from any of the 7 X-ray detection layers. This condition is relaxed if there is a simultaneous second pulse within the K-band i.e. events at about 29.8 keV due to detection of fluorescent X-rays of xenon in any of the layers. Broad band counting and pulse height analysis of all the valid X-ray events are done using 12 bit counters and a 1024 channel pulse height analyser. The total processing time (dead time) for analysed events about 32 $\mu$s. A detailed description of the signal processing electronics will be presented in a separate paper.

\subsubsection{System Time Base Generator (STBG)}
This system provides a stable and accurate timing reference of 10 $\mu$s accuracy for the LAXPC electronic sub-systems as well as time reference for other X-ray payloads on the ASTROSAT mission. It provides command selectable time bins, from 8 ms to 1024 ms for broadband counting and from 1 seconds to 128 seconds  for pulse height histograms of post-processing electronics. It also provides 32-bit one-second counter, which will provide a continuous time reference throughout the mission life.  It uses highly stable low drift oven controlled crystal oscillator of 10 MHz frequency with $\pm$ 0.2 PPM stability of frequency in -20 to +70$^{\circ}$C range.

\subsection{The LAXPC Gondola and Source Pointing and Tracking System}
The two LAXPCs are mounted in a sturdy cradle (Figure \ref{fig3}) that is driven in elevation by a geared DC motor with 10 bit shaft encoder to measure the elevation angle. The detector cradle is mounted on an oriented platform which is coupled at the top to a servo control system consisting of a reaction wheel driven by a DC torque motor. This in turn couples the payload with the balloon. A photograph of the assembled LAXPC instrument before the balloon flight is shown in Figure \ref{fig3}.

\begin{figure}
\includegraphics[width=8cm]{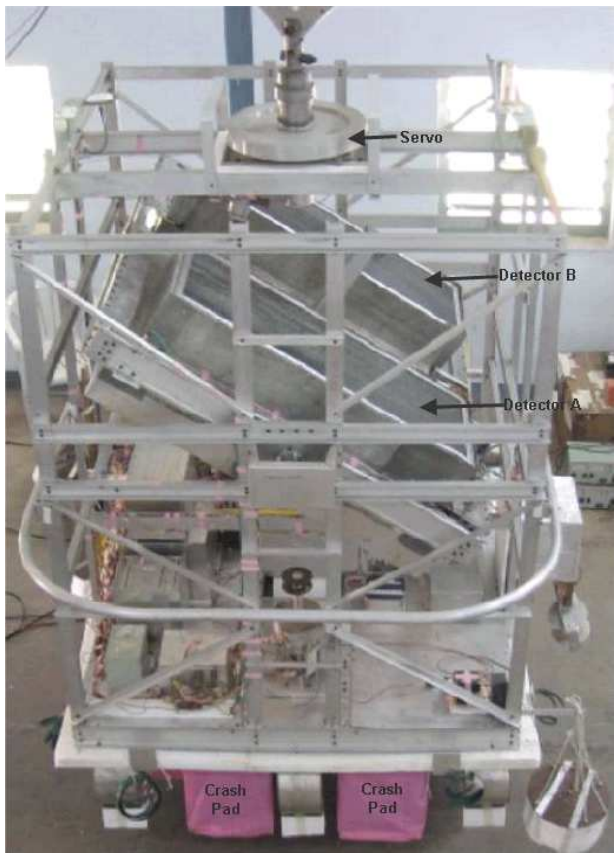}
\caption{ A photograph of the balloon instrument ready for launch, with the 2 LAXPC detectors mounted in a cradle and various subsystems on the oriented platform.}
\label{fig3}      
\end{figure}

The azimuth angle of the payload can be controlled and rotational perturbations due to balloons are corrected by the servo system. A flux gate sensor mounted on a small platform driven by a geared motor, is used as a sensor for the magnetic field of the Earth to derive error signal from the null position for controlling the azimuth orientation. The azimuth angle is measured by a 10 bit shaft encoder. The movements in azimuth and elevation directions are controlled by a source tracking program loaded in an onboard memory with the provision of intervention by radio command. This permits acquisition, pointing and tracking of any source whose coordinates (right ascension and declination) are stored in the onboard memory.  The oriented instrument is automatically moved in elevation and azimuth every minute for tracking a source. The azimuth and zenith angles of different X-ray sources are calculated onboard for the flight on a given day. The coordinates are converted into corresponding shaft encoder readings and stored in an on-board EPROM. These stored values of azimuth and elevation angles are compared during the flight with the actual readings to activate the motors and correct the difference. The system start time is clocked by an alarm time which is set on the ground just before the launch of the payload. This method provides the on-board tracking of a given X-ray source during the flight. In case of any malfunction of the program, the source tracking can also be done by sending manually real time telecommands. A source can be tracked to an accuracy of about 0.5$^{\circ}$ by this method.

\subsection{Measuring Aspect of the Instrument}
A crossed pair of flux gate sensors (called X and Y) are used to measure the actual orientation of the payload in azimuth and elevation. For this purpose the X and Y flux gate sensors are calibrated on the ground before the flight and also during the flight by changing the azimuth and elevation angles by telecommand in discrete steps. Calibration of the flux data sensors during the flight ensures confidence in the correct pointing and deducing the aspect information. 

\section{Telecommand and Telemetry of Data}
Radio Frequency link operating in S band is used between the air-borne payload and the ground station for control of the LAXPC instrument and its operation as well as flight support subsystems. This is done by a telecommand system consisting of (i) an encoder on the ground that encodes the information and transmits it to the payload by a 2080 MHz transmitter, installed at the ground station, (ii) an onboard receiver and decoder unit to receive and decode the information received from the ground station and sends it to the proper on-board units for further controls. 
The transmission of the data from the instrument to the ground is through a S-band (2.2 GHz) telemetry link. The Telemetry Encoder is an onboard fully configurable system designed for encoding the instrument data, house-keeping data and health parameters of the instrument. Digital data stream conversion is done from Non-return to zero (NRZ) code to Bi-phase level (Bi-$\Phi$-L) code. This signal is further routed via a high-pass filter to make it bipolar and fed to the transmitter for modulation and transmission.

\subsection{Telemetry Format  and Different Type of Events}
The flight parameters, housekeeping data, and detector science observation data, are transmitted in a predetermined frame format at telemetry bit rate of 100 kHz. All the data are arranged in sub-frame format consisting of 231 sub-frames of 54 bytes/words each. Data contents of different sub-frames of LAXPC are summarized in Table \ref{Sub-Frame-Description.tab}. 

\begin{table}
\begin{center}
\caption{Telemetry Sub-frame Description of LAXPC}
\label{Sub-Frame-Description.tab}
\begin{tabular}{ll}
\hline
Frames & Description \\
\hline\\
105 & Detector A events\\
105 & Detector B events\\
16 & Anti-spectrum information\\
3 & Housekeeping Information\\
2 & Counter Information\\[6pt]
\hline
\end{tabular}
\end{center}
\end{table}

The following data are transmitted in the telemetry format per sub-frame: 
1. Timing and energy information of the events from the 2 detectors are captured.\\
2. Veto layer data for each detector are transmitted.\\
3. Housekeeping information like HV values, various voltages, temperatures of different parts of the instrument, azimuth and elevation data, aspect  information etc are transmitted in 3 sub-frames.\\
4. Digital Event Counters like Integral counts in 3-80 keV (INTCC), counts of events exceeding the upper energy level of 80 keV (ULD), counts of all events triggering the lower energy threshold of 3 keV (LLD), Veto layer count rates (ANTI), number of Xenon-K fluorescent events detected (NK) for each detector, Checksum bytes for error checking in each frame etc. In the LAXPC balloon flight telemetry system, data is transmitted in bytes. Hence there is a requirement of sync byte pattern to know the start of each frame. For LAXPC balloon data frame "4D 78" (Hex format) are sync bytes which indicate start of a frame.
Details of different types of events stored in the Broad-band counters and from the event processing logic system of each LAXPC are summarized in Table \ref{Eventprocessinglogic.tab}.\\

\begin{table}
\begin{center}
\caption{Details of events stored in Broad-band Counters of LAXPC Event processing logic (EPL system).}
\label{Eventprocessinglogic.tab}
\footnotesize
\begin{tabular}{lll}
\hline         
No.	&Counters	&Description of events\\
\hline \\
1	&LLD		&Event Crossing the lower energy threshold of 3 keV in any of \\
	&		&the Main Anodes A1 to A7\\
2	&ULD		&Event Crossing the upper energy threshold of 80 keV in any of\\
	&		&the Main Anodes A1 to A7\\
3	&ANTI		&Event in any of three Anti Anodes A8 to A10 of with threshold $>$ 3 keV\\
4	&MTO		&More than One Simultaneous Event within 4$\mu$s in any of 10 anodes\\
5	&MTT		&More than Two Simultaneous Event within 4$\mu$s in any of 10 anodes\\
6	&ANTI-1		&Event of $>$ 3 keV in Right Anti Anodes (A8)\\
7	&ANTI-2		&Event of $>$ 3 keV in Left Anti Anodes (A9)\\
8	&ANTI-3		&Event of $>$ 3 keV in Bottom Anti Anodes (A10)\\
9	&NK		&MTO (more than one Event) one of which falls in K-band (25-35 keV)\\
        &               &and is taken as due to detection of Xenon-K fluorescent photon.\\ 
10	&3-80 keV	&Integral counts of all valid X-ray events in 3-80 keV band\\
11	&3-6 keV	&Non-MTO Events between 3-6 keV in Particular Layer / Anode\\
12	&6-18 keV	&Non-MTO Events between 6-18 keV in Particular Layer / Anode\\
13	&18-40 keV	&Non-MTO Events between 18-40 keV in Particular Layer / Anode\\
14	&40-80 keV 	&Non-MTO Events between 40-80 keV in Particular Layer / Anode\\
15	&INTCC		&Integral count rate of pulse height analysed events \\[10pt]
\hline
\end{tabular}
\end{center}
\end{table}

Event Information encoded in 8 bytes/words of each frame is as follows: \\
W4 = Detector ID; (DA or DB) \\
W5 = T3 (Time tag) \\
W6 = T2 (Time tag) \\
W7 = T1 (Time tag) \\
W8 = Anode-ID (0-7 anodes)\\ 
W9 = PHA byte of 1st event \\
W10 = PHA byte of 2nd event \\
W11 = PHA (4-bits each) of 1st and 2nd event \\
The information is computed as following:  Event time = T3+T2+T1 ($\mu$s), Pulse Height Amplitude 1st event = W9 + W11 (4 lower bits), Pulse Height Amplitude 2nd event = W10 + W11 (4 upper bits).

The telemetered data are fed to a multi-channel commercial digital telemetry processing unit (D/PAD). This unit performs bit, frame and sub-frame synchronizations and de-commutation. It displays 12 selected words as well as provides output words along with analogue outputs for these words. Out of these 8 outputs are displayed on strip chart recorders for monitoring the working of the instrument. It is then passed to a Decoder and PC unit through a bit synchronizer and a frame synchronizer to extract information on science data such as count rates, pulse height spectrum of detected events, various housekeeping data and parameters, aspect information etc. All the data are decoded on-line and displayed in real time on two PC's. 

\section{LAXPC Calibration, Flight Performance and Data Analysis}
A pre-flight calibration of both the detectors was carried out using a collimated Americium 241 (Am-241) radioactive source which emits 59.6 keV mono energetic X-rays. The LAXPC detectors were filled with Xenon (90\%) and Methane gas (10\%), at a pressure of 1508 torr in detector A and 1472 torr in detector B. Appropriate HV was applied to each detector. The detectors were calibrated by placing the collimated source on top of the FOV collimator. The average response of each detector was obtained by shining the source for equal duration of time at different equally spaced locations all over the detector. Pulse height spectrum was acquired at each location with a Multi Channel Analyzer (MCA) and all the pulse height spectra were summed to obtain the average spectral response of the detector. Background count rates were measured for 3000 sec for each of the two detectors after removing the Am-241 source. 
The background subtracted pulse height distribution plots obtained from the two detectors, LAXPC A and LAXPC B are shown in Figure \ref{fig4}. Each plot shows two pulse height peaks, one at 29.8 keV, due to escape of 29.8 keV K-fluorescent photon of Xenon and second at 59.6 keV, being the full photo-peak of the X-ray line from Am-241. Two peaks were observed at MCA channels 230 and 439 for detector A and channels 237 and 465 for detector B. For LAXPC A, the background subtracted energy resolution defined by FWHM at 29.8 keV and 59.6 keV were measured to be 23\% and 26\% respectively. For LAXPC B, the corresponding resolutions were measured to be 20.6\% and 16.8\%. The rather poor energy resolution was due to prolonged exposure of the detectors to the atmosphere due to which water vapor and oxygen molecules diffused through the Mylar window and degraded the energy resolution. Due to technical difficulties it was not possible to purify the Xenon gas in the LAXPCs just before the balloon flight.

\begin{figure*}
\resizebox{\hsize}{!}
{
\includegraphics*[width=9cm,angle=270]{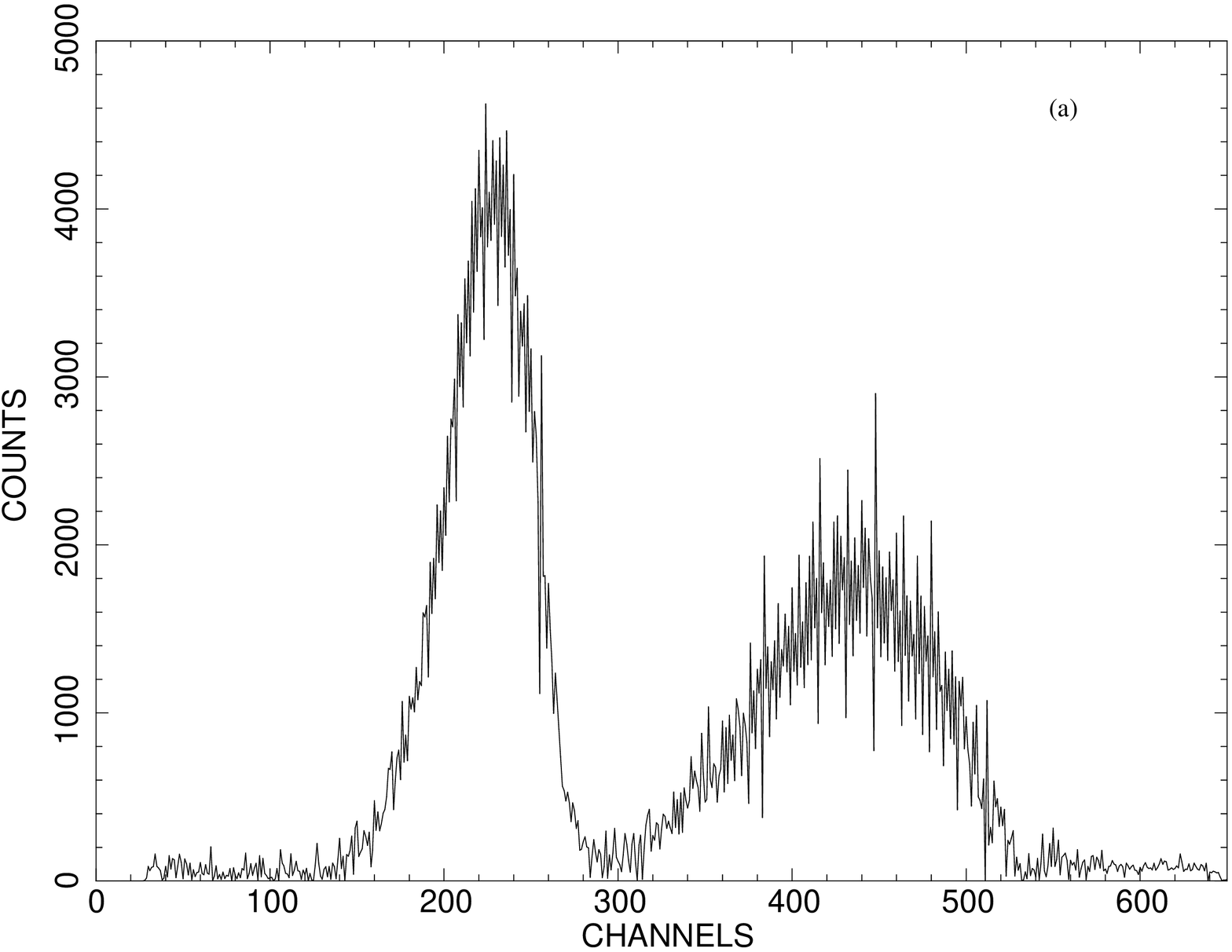}\hspace*{0.5em}
\includegraphics*[width=9cm,angle=270]{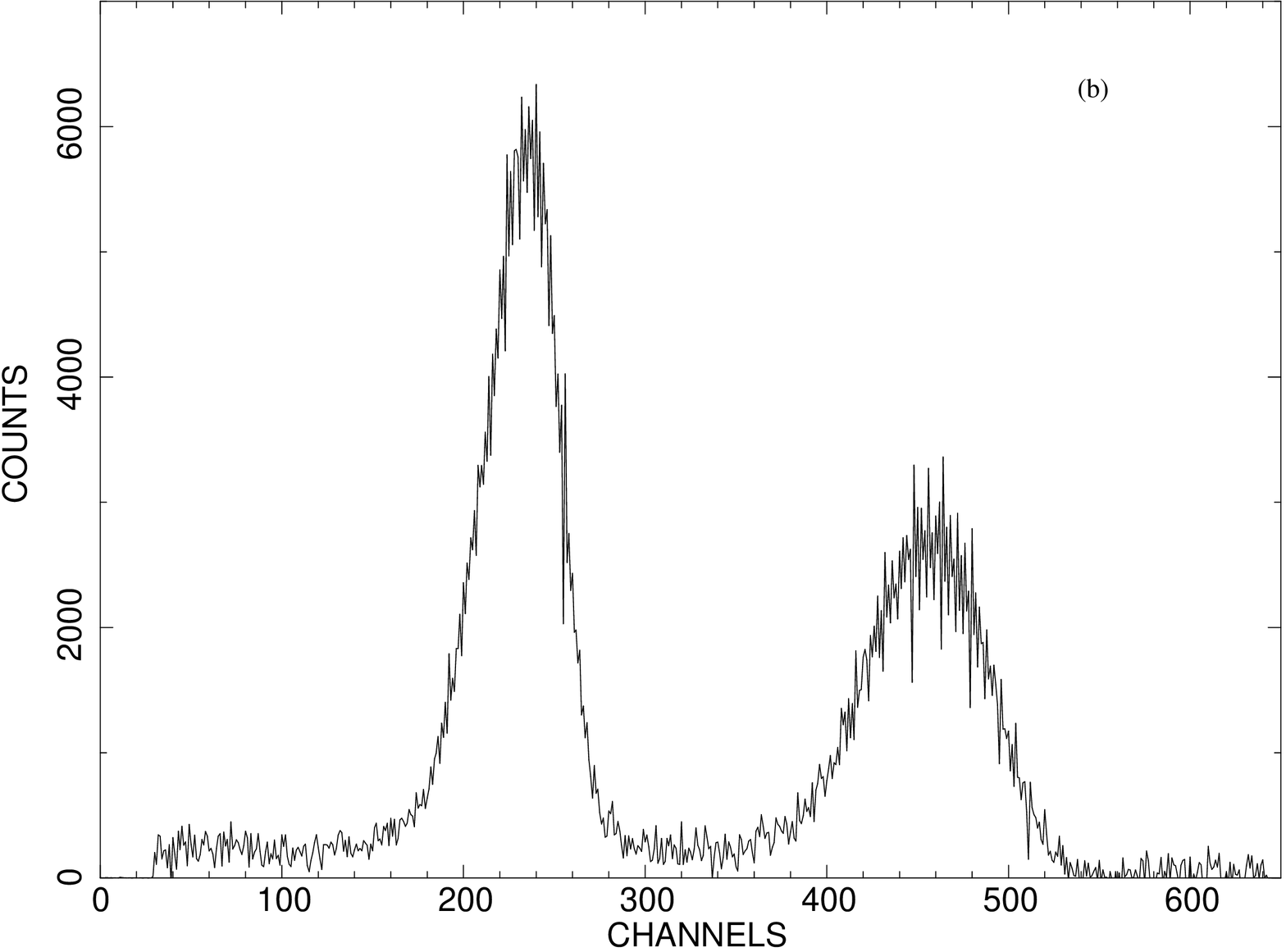}\hspace*{0.5em}}\\
\caption{Energy resolution of LAXPC A and B measured by averaging the response over the entire detector area using a collimated Am-241 source. The peaks at 29.8 keV and 59.6 keV are shown in the figures (a) For LAXPC detector A and (b) For LAXPC Detector B}
\label{fig4}
\end{figure*}

The LAXPC instrument was launched on a 26 million cubic feet volume balloon on 2008 April 14 at 12.40 hours.  It reached ceiling altitude of 41 km corresponding to a vertical residual atmosphere of 2.5 g/cm$^2$ at 2.40 hours.  The balloon floated at the ceiling for 6 hours and 36 minutes. Balloon flight was terminated at 9 hours 16 minutes on 2008 April 14, by telecommand. The main objective of the flight was to assess the performance of LAXPCs by making measurements of the background as well as  pointed mode observations of X-ray binary Cyg X-1, the brightest X-ray source above 20 keV at balloon altitudes. The observation program included pointed mode Cyg X-1 data and measurements of the background by pointing away from the source before and after Cyg X-1 observation. These data was used for assessing the performance of LAXPC detectors and analysing time variability of Cyg X-1. 
After the balloon reached the ceiling altitude a source-free region in the sky was observed to measure the background count rates. Following this Cyg X-1 was observed for 3 hours 19 minutes. The tracking was so programmed that observations of the source-free background region and Cyg X-1 are carried out alternately during the flight. The source and background regions were observed alternately to correct for any systematic variations in the background count rates. Cyg X-1 data were acquired from both the LAXPC detectors. Detailed log of the X-ray observations made by the two LAXPC detectors A and B is given in Table \ref{ObservationLog.tab}.

\begin{table}
\begin{center}
\caption{Log of X-ray observation of Cyg X-1 and background with LAXPC payload on 2008 April 14.}
\label{ObservationLog.tab}
\small{
\begin{tabular}{llll}
\hline        
S.No& Pointed Observation & Time of Observation & MJD\\
\hline\\
1.& Time of launch & 12.40 am & 54570.0277\\
2.& Time to reach ceiling & 2.40 am& 54570.1111\\
3.& Background (BI) & 2.45-2.55 am& 54570.1145-54570.1215\\  
4.& Background (BII)& 3.56-4.05 am& 54570.1638-54570.1701\\ 
5.& Cygnus X-1 & 4.06-6.15 am& 54570.1708-54570.2604\\
6.& Cygnus X-1 & 6.26-7.25 am& 54570.2680-54570.3090\\
7.& Background (B10 III)& 7.26-7.35 am & 54570.3097-54570.3159\\
8.& Balloon flight termination  & 9.16 am & 54570.3861\\
   &(Main cut) &&\\[10pt]	
\hline
\end{tabular}
}
\end{center}
\end{table}

Raw data from different layers, anodes and energy bands (listed in Table \ref{Eventprocessinglogic.tab}) were extracted using LabVIEW software in hex format which was then converted to ASCII. These files were subsequently made into fits file format which is compatible with Heasarc  XRONOS software \cite{Stella 1992}. The ASCII files were converted into spectrum files with respect to different channels using ascii2pha command in XSPEC package \cite{Arnaud 1996}. Information of pointing vector is stored in the house keeping data obtained from the detector during the source and background observations. Using this information we calculated the offset of detector pointing direction from the source position. This was then used to compute the source flux by applying proper offset correction. The offset corrected Cyg X-1 count rates were derived by dividing the observed count rate with the exposure efficiency. 
Light curves were constructed for the entire flight, using 1 second time resolution 3-80 keV integral mode data for both the detectors and these are shown in Figure \ref{fig5}. 

\begin{figure*}
\resizebox{\hsize}{!}
{
\includegraphics*[width=8cm, angle=-90]{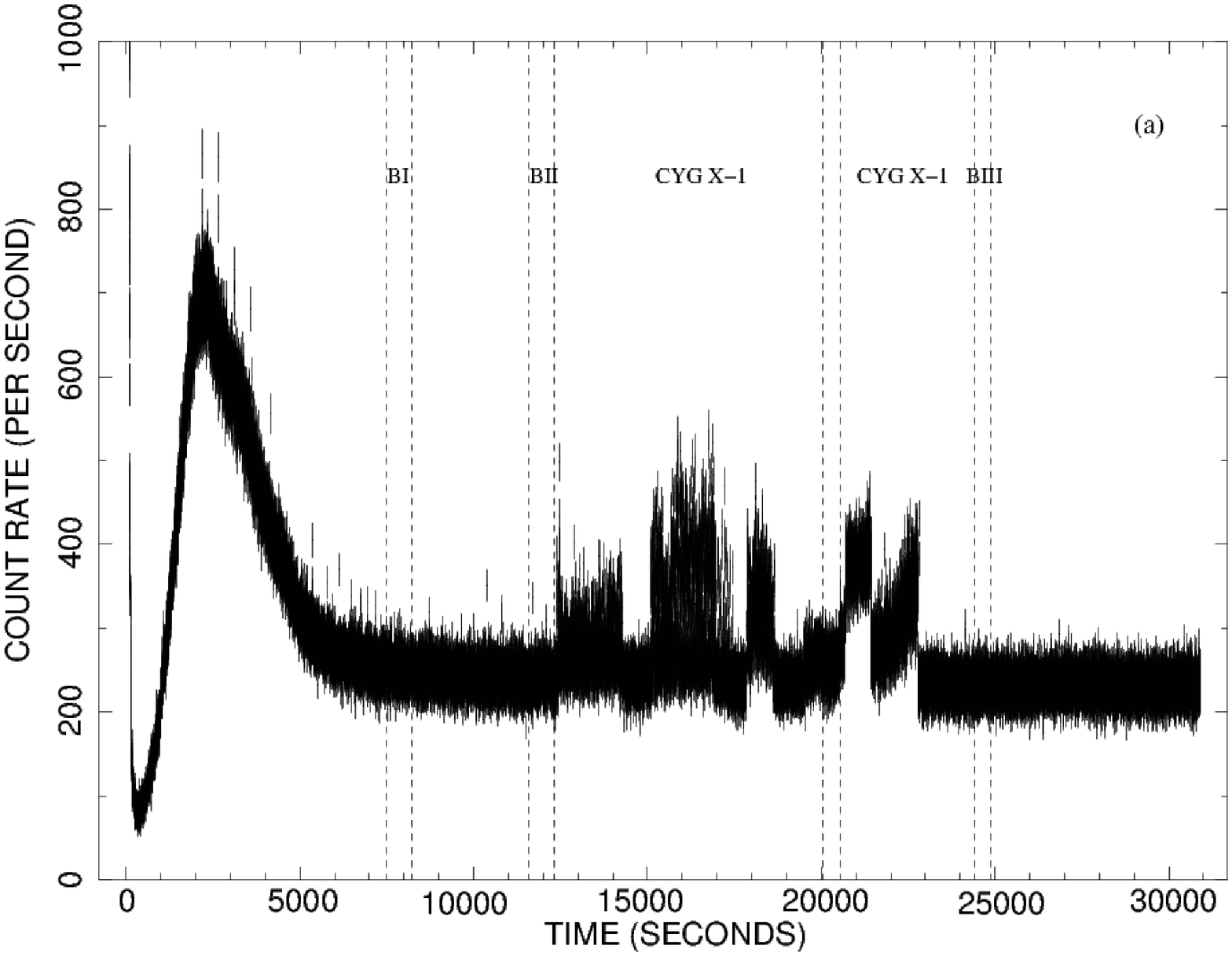}\hspace*{0.5em}
\includegraphics*[width=8cm, angle=-90]{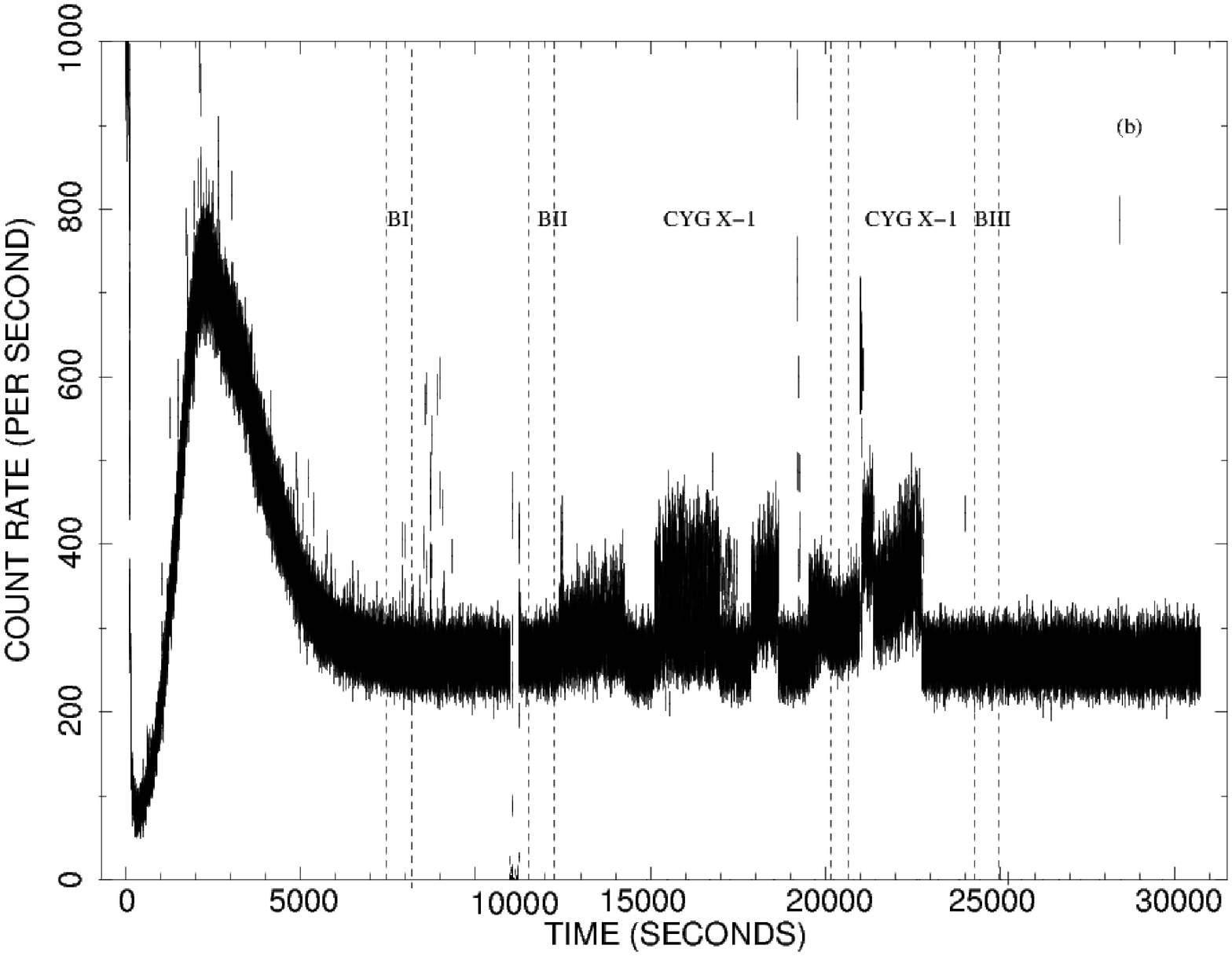}\hspace*{0.5em}}\\
\caption{Light curve generated from LAXPC Integral count rates in 3-80 keV obtained for the entire duration of the flight starting from MJD 54570.0277 (12.40 am on 2008 April 14) are presented for LAXPC A in Panel (a) and  LAXPC B in Panel (b). Stretches of Cyg X-1 observation (from 4:06-6.15 am and 6.26-7.25 am on 2008 April 14, seen as increased count rates) and background observations (BI, BII and BIII) are indicated in the plot presented in the figure.}
\label{fig5}
\end{figure*}

\section{Results}
Except for brief intervals when the HV of Detector B fluctuated, both the detectors functioned well during the experiment and provided useful data. In detector B, the HV was switched off manually at 2 hours 43 minutes after the launch of the flight for 5 minutes as intermittent HV discharge was noticed. Pressure, high voltage and temperature were continuously monitored for both the detectors. These parameters were stored in the house keeping data.

Count rates for the individual layers of the detectors were extracted in different energy bands using broad band counter. These count rates (counts per minute) for each layer as well as the summed count rates for all the layers of each detector in different energy intervals are used to construct light curves from the time the balloon reached the ceiling altitude till the end of the balloon flight.  For background estimation, we selected the background observation stretches BI, BII and B III indicated in Figure \ref{fig5}. Pulse height array mode data were extracted for the entire flight duration, including Cyg X-1 and background observations. 
During the pointed mode observations of the background and the source, it was noticed that at some times the gondola did not stabilize in the pointing direction but was rather oscillating around the pointing direction with a period of about 30 sec. Therefore data from only those observations were analysed where gondola orientation was found to be stable. Integral count rates with one second time resolution were extracted to check the performance of both the detector during the flight. Average background count rates of both the detectors derived from integral 3-80 keV data were found to be constant for both the detectors. Background count rates in different energy bands were also obtained from broad band counters for both the detectors. 

\subsection{Count rates in Different Energy Bands of the Detectors}
To check the consistency of performance of the LAXPC detector, layer wise (A, B and C) broad band count rates (counts per minute) are plotted in energy bands 3-6, 6-18, 18-40, and 40-80 keV and in 3-80 keV band for the LAXPC detectors A and B. It should be pointed out that the broad band count rates include only Non-MTO events and therefore exclude all those events in which 29.8 keV Xenon K-fluorescent photon is simultaneously detected in another anode and therefore registered as MTO event. On the other hand the integral 3-80 keV count rates include all the accepted X-ray events that are analysed for pulse height. This includes MTO events, one of which is due to detection of 29.8 keV Xenon K-fluorescent photon in a different layer. The Integral 3-80 keV count rate will therefore be very substantially higher compared to the broad band count rates. The count rate plots for detector A are shown in Figure \ref{fig6}, Figure \ref{fig7}, and Figure \ref{fig8}. Similar results and count rate variabilities are observed from detector B. We have also plotted the summed count rates from all the layers of the detector A and B and these are shown in Figure \ref{fig9} and Figure \ref{fig10} in different energy bands. Count rates of background and Cyg X-1 observations derived from broad band counters in different energy intervals for Detector A are tabulated in Table \ref{countrates.tab}. These light curves were checked for rapid timing variability.

\begin{figure}
\centerline{
\includegraphics*[width=8.0cm, angle=270]{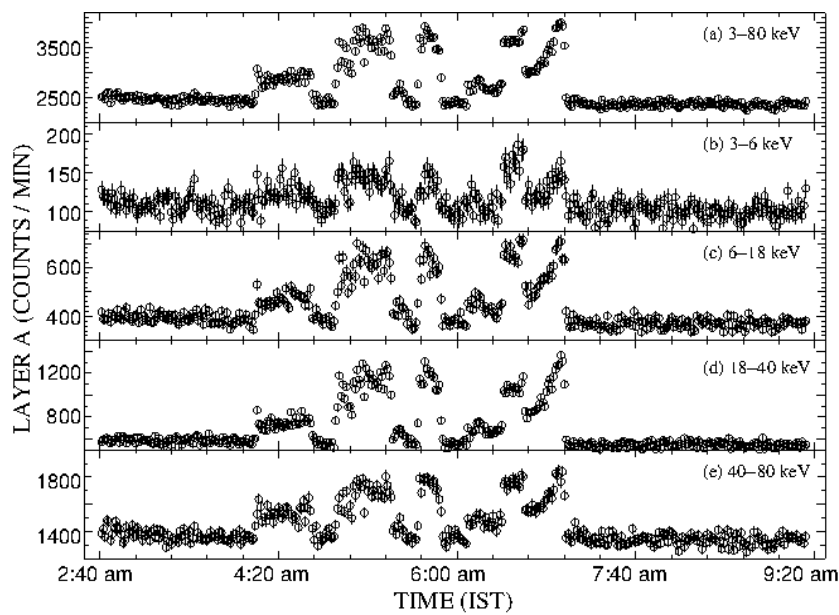}}
\caption{Light curves obtained from layer A using Broad Band Counters data in different energy bands for
 Detector A after the balloon reached ceiling till the end of the flight. Layer A is summed output of odd and even anodes in top layer i.e. layer 1.}
\label{fig6}
\end{figure}

\begin{figure}
\centerline{
\includegraphics*[width=8.0cm, angle=270]{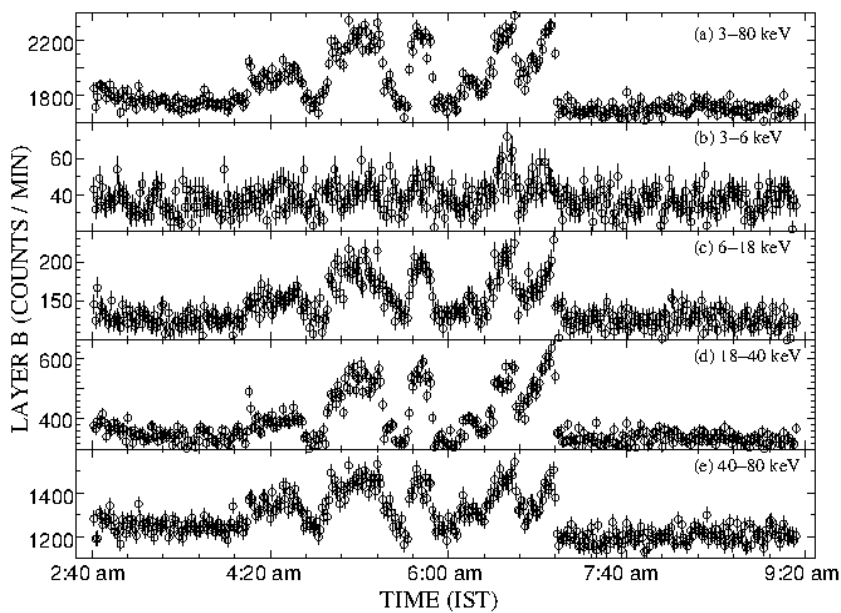}}
\caption{Light curves generated from layer B using Broad Band Counters data in different energy bands for
 Detector A after the balloon reached ceiling till the end of the flight. Layer B is summed output of odd and even anodes in middle layer i.e. layer 2.}
\label{fig7}
\end{figure}

\begin{figure}
\centerline{
\includegraphics*[width=8.0cm, angle=270]{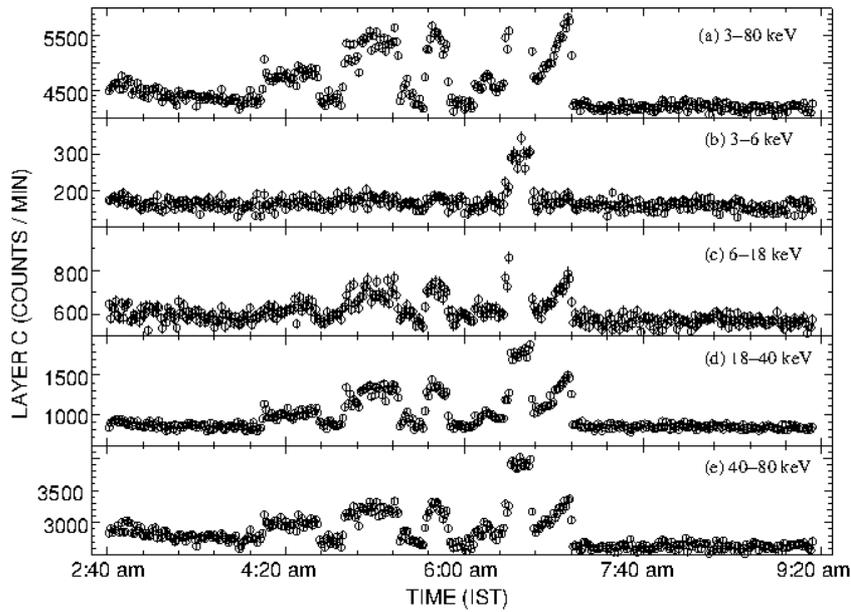}}
\caption{Light curves generated from layer C using Broad Band Counters data in different energy bands for
detector A after the balloon reached ceiling till the end of the flight. Layer C represents the combined output of layers 3, 4 and 5.}
\label{fig8}
\end{figure}

\begin{figure}
\centerline{
\includegraphics*[width=8.0cm,angle=270]{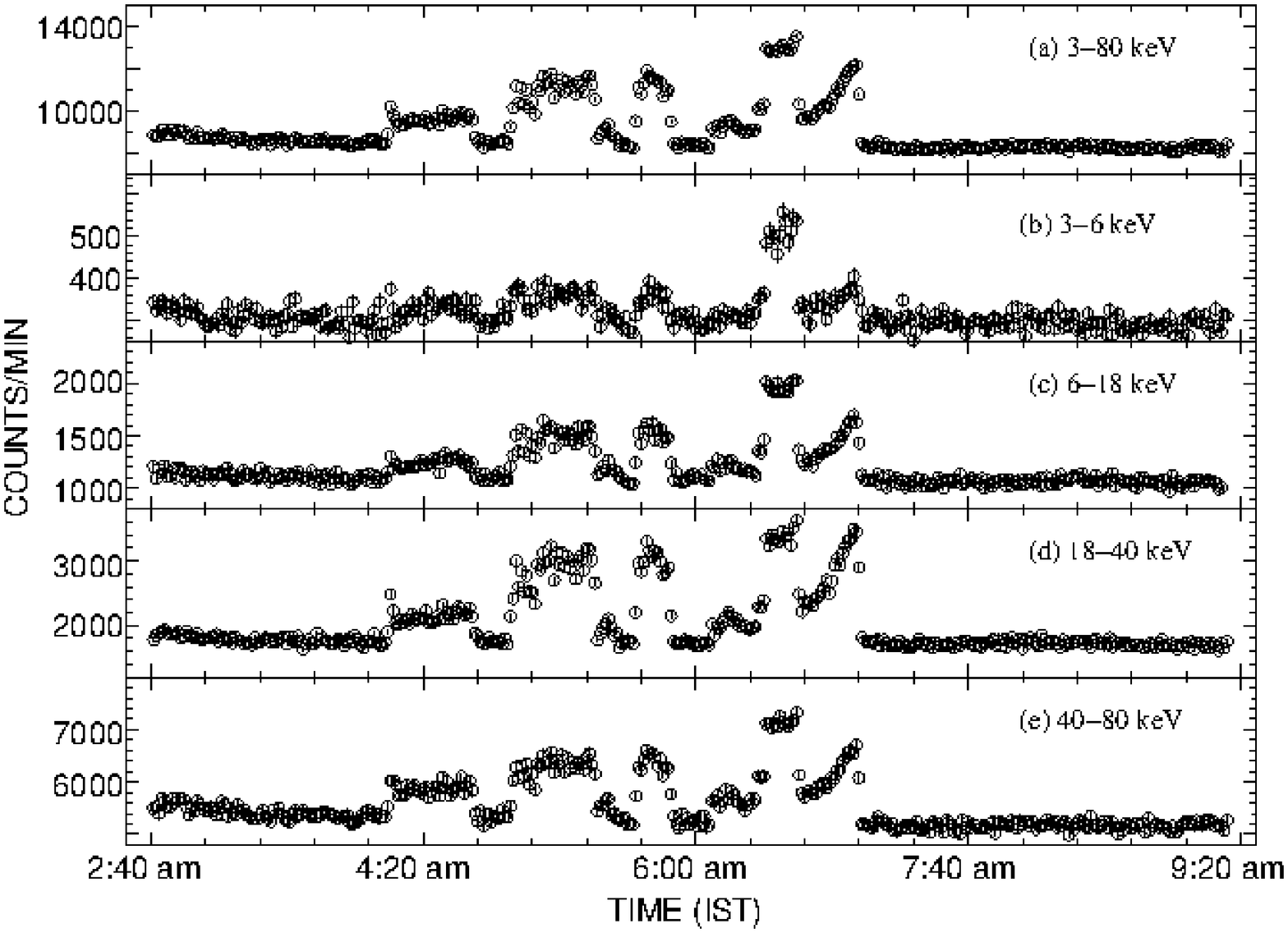}}
\caption{Light curves generated from summed Broad Band count rates of all the layers from Detector A in different energy bands after the balloon reached ceiling till the end of the flight.}
\label{fig9}
\end{figure}

\begin{figure}
\centerline{
\includegraphics*[width=8.0cm, angle=270]{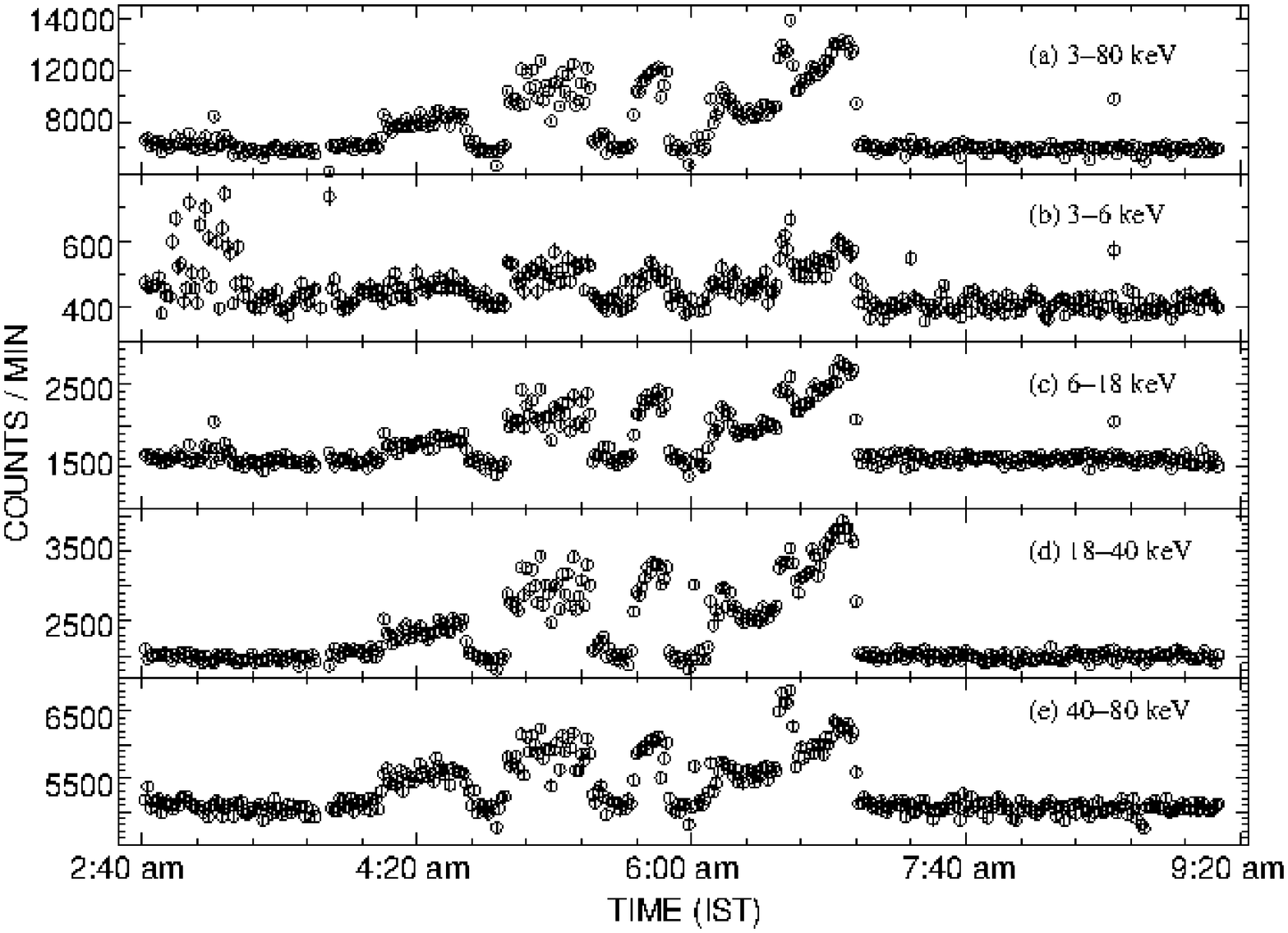}}
\caption{Light curves generated from summed Broad Band count rates of all the layers from Detector B in different energy bands after the balloon reached ceiling till the end of the flight.}
\label{fig10}
\end{figure}

\begin{table}
\begin{center}
\caption{Count rates of background and Cyg X-1 observations derived from Broad Band Counters in different energy intervals for Detector A.}
\label{countrates.tab}
\small{
%\begin{tabular}{>{\bfseries}l>{\bfseries}l>{\bfseries}l>{\bfseries}l>{\bfseries}l>{\bfseries}l}
\begin{tabular}{llllll}
\hline        
& \multicolumn{5}{c}{ENERGY BANDS ($\times$10$^{-4}$ counts/ cm$^{2}$ s keV)}\\
LAXPC A&&&&& \\
	&3-6 keV & 6-18 keV & 18-40 keV & 40-80 keV & 3-80 keV\\
BACKGROUND&&&&& \\
LAYER A&2.3$\pm$0.04 &2.0$\pm$0.02&1.6$\pm$0.01&2.1$\pm$0.01&2.0$\pm$0.008\\

LAYER B&0.8$\pm$0.03 &0.7$\pm$0.01&1.0$\pm$0.01&1.9$\pm$0.01&1.4$\pm$0.007\\

LAYER C&3.4$\pm$0.05 &3.1$\pm$0.03& 2.4$\pm$0.007&4.4$\pm$0.02&3.6$\pm$0.01\\

 &&&&&\\

SOURCE+BACKGROUND &&&&&\\
LAYER A&2.6$\pm$0.01 &2.7$\pm$0.01&2.3$\pm$0.001 &2.4 $\pm$0.005 &2.5$\pm$0.003\\

LAYER B&0.9$\pm$0.01 &0.8$\pm$0.01&1.2$\pm$0.005&2.1 $\pm$0.004 &1.6$\pm$0.003\\

LAYER C&3.7$\pm$0.02 &3.5$\pm$0.01&3.1$\pm$0.005&4.7 $\pm$0.007 &4.0$\pm$0.004\\
  
\hline
\end{tabular}
}
\end{center}
\end{table}

\begin{figure}
\centerline{
\includegraphics*[width=8.0cm, angle=270]{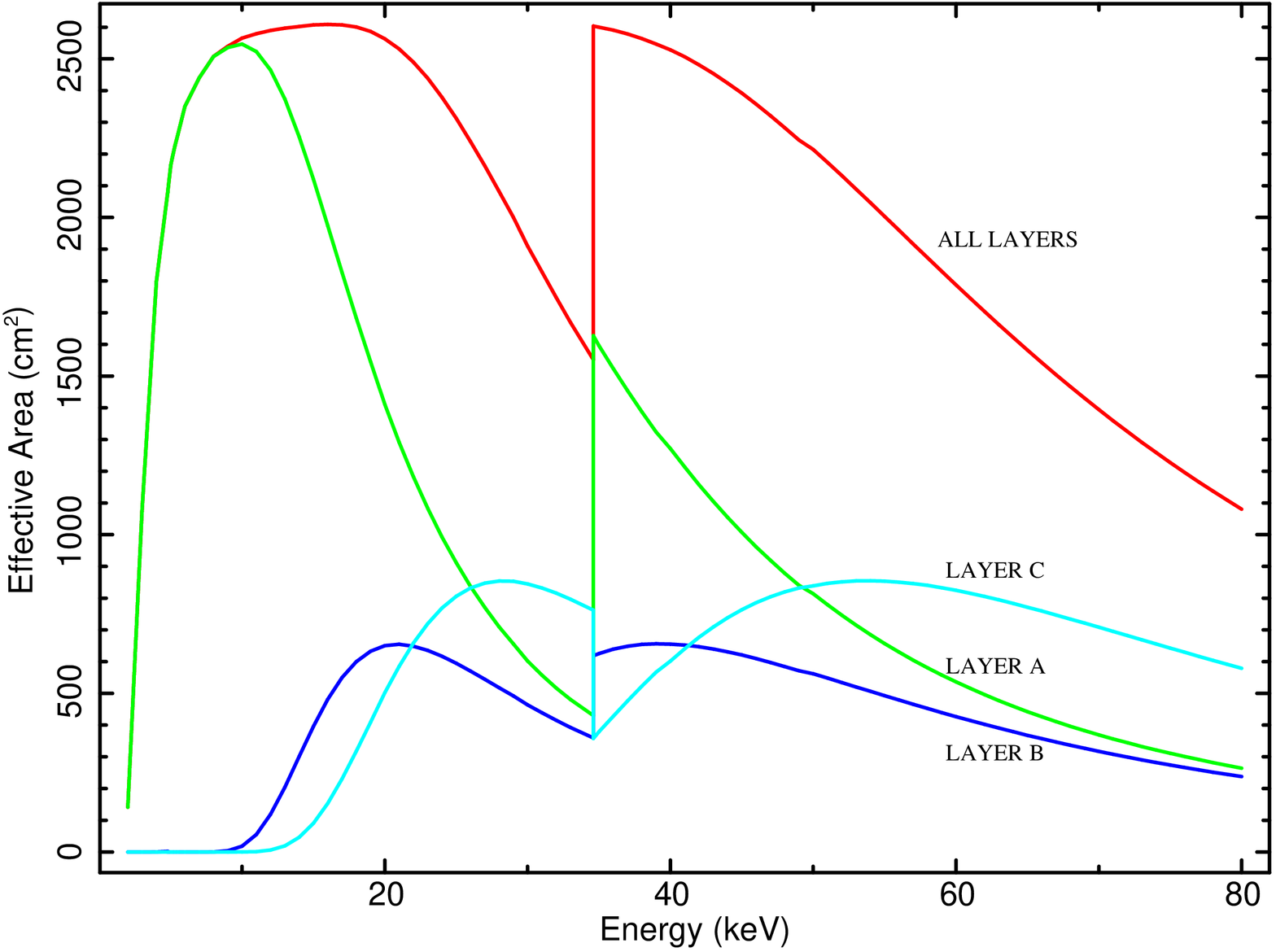}}
\caption{Layer wise and summed effective area plot of LAXPC A.}
\label{fig11}
\end{figure}
\subsection{Detection Efficiency of LAXPCs}
To understand the count rates in the individual layers it is necessary to compute detection efficiency of each layer of LAXPCs. Using photon absorption cross sections from \cite{Veigele 1973} the detection efficiency of each layer of Detector A was computed for photons in 3-80 keV range incident on the detector. Attenuation in Mylar window was folded in the computed values. This efficiency was converted layer wise into effective detector area as a function of photon energy. A geometrical area of 3600 cm$^{2}$ and collimator transmission of 0.73 was used in this computation of effective area of each detector.

 A plot of detection efficiency for individual layers A, B and C as well as the summed efficiency of all the layers for Detector A is shown in Figure \ref{fig11}. Similar efficiency curve is also valid for Detector B. Below 10 keV almost all the incident photons get absorbed in Layer A. Between 10 and 20 keV also the maximum number of photons interact in Layer A followed by Layer B and C. Between 20 and 34.5 keV (xenon K-edge) the interaction probability drops rapidly in all the Layers and then increases again above Xenon K-edge. It will be noticed from the Figure \ref{fig11} that for photons of less than 50 keV maximum detection occurs in Layer A followed by Layer C and then Layer B. At 50 keV and higher energy maximum detection efficiency is for Layer C due to its large interaction depth of 9 cm.

The maximum count rate is accordingly detected from Layer C which consists of combined outputs of three anodes 5, 6 and 7. The second highest count rate during the balloon flight was observed from the top layer A in which X-rays entering from the collimator undergo first interaction. Since photons of $<$ 50 keV get significantly absorbed in the layer A and only remaining X-rays enter the layer B with depth of 3 cm, the count rate in the layer B is lower. The same behavior of X-ray count rates in different layers is observed from both the detectors. It should be pointed out that at a float altitude of about 2.5 g/cm$^{2}$ of residual atmosphere, only energetic X-rays of $\ge$ 15 keV and above can reach the detectors from an extra-terrestrial source. The count rate enhancements observed in Figures \ref{fig6} to \ref{fig10} in 3-6 and 6-18 keV bands are either due to atmospheric X-ray background and internal detector background due to cosmic rays interactions or due to ''Escape Effect'' of X-rays having energy greater than 34.5 keV i.e. K-edge of Xenon. When photons of $>$ 34.5 keV interact in the xenon gas in about 90\% of interactions a 29.8 keV fluorescent X-ray of xenon is produced which may not interact in the same layer but may be absorbed in any other adjacent layer or may altogether escape from the detector without interaction. Difference between the energy of the incident X-ray and that of the fluorescent photon is deposited in the layer in which interaction takes place. Significant count rates seen in Figure \ref{fig6} to \ref{fig10} below 20 keV, in the background as well as Cyg X-1 observations arise due to the ''Escape'' effect discussed above. Light curves obtained from LAXPC Integral data in one second bins were obtained during the 4.06-6.15 am and 6.26-7.25 am stretch of Cyg X-1 observation on 2008 April 14. Background subtracted count rates (counts/second) of Cyg X-1, with counts averaged over 10 seconds from both stretches of observations from both the detectors A and B, are shown in Figure \ref{fig12}.

\begin{figure}
\centerline{
\includegraphics*[width=5cm, angle=-90]{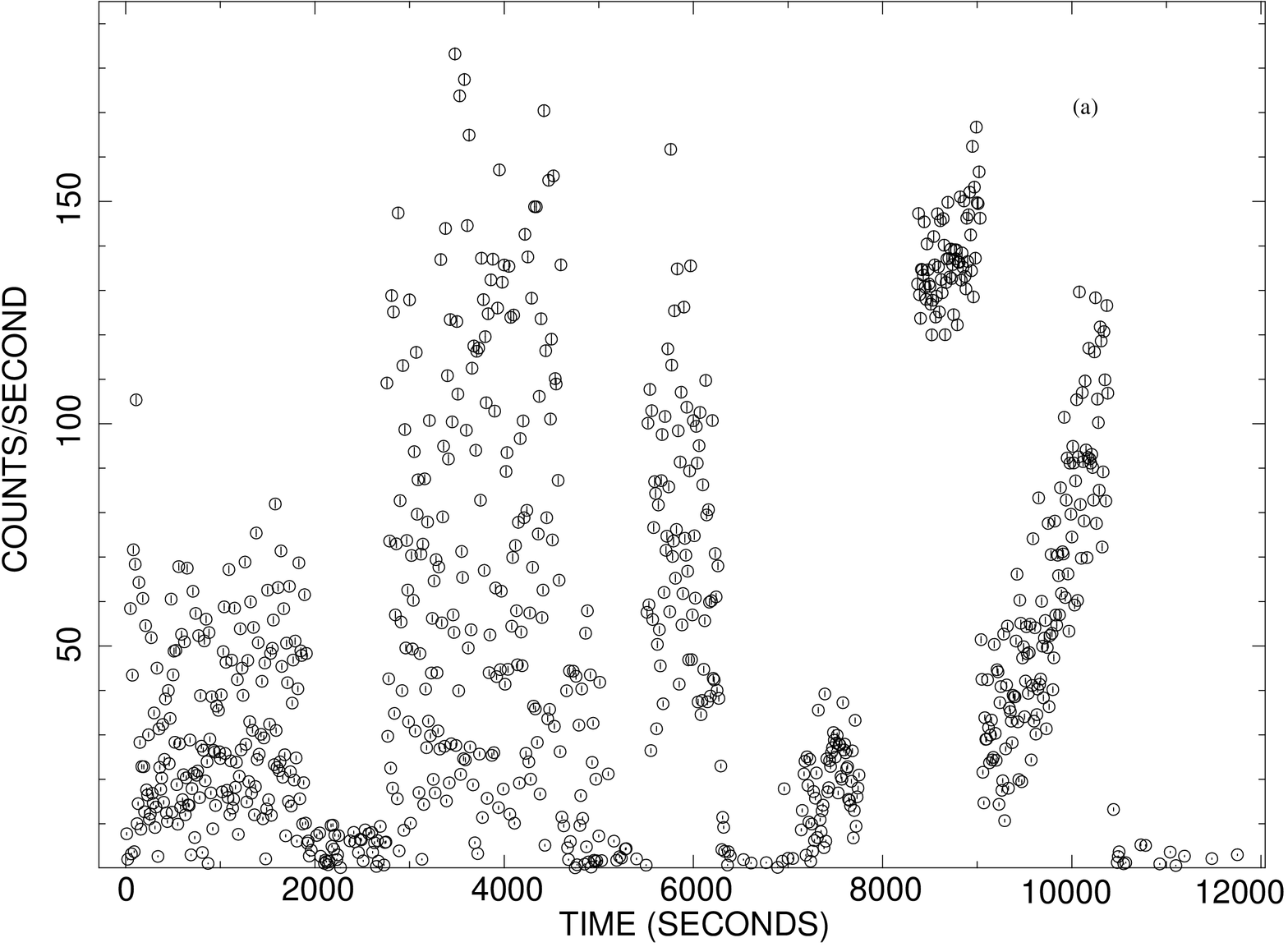}\hspace*{0.5em}
\includegraphics*[width=5cm, angle=-90]{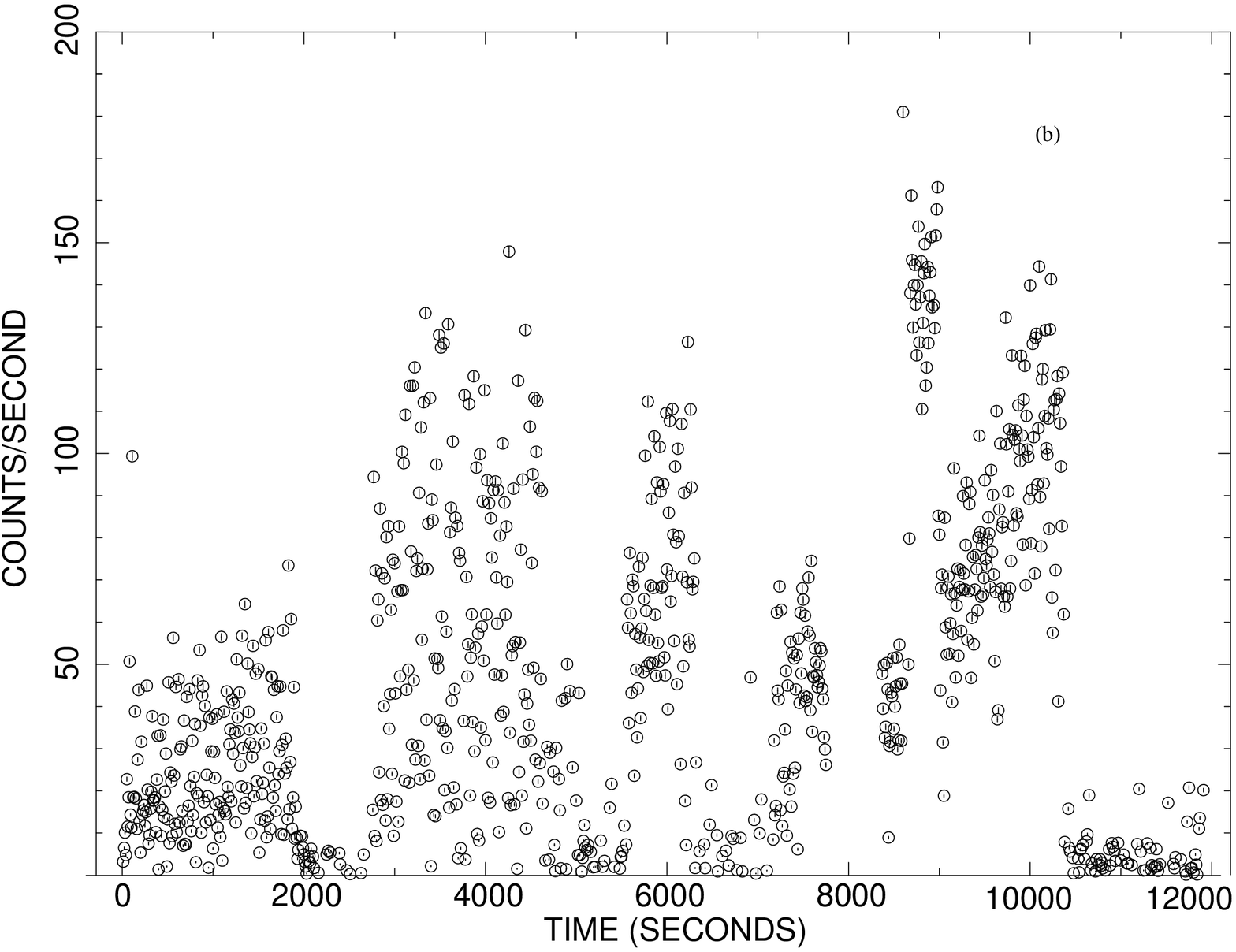}\hspace*{0.5em}}
\caption{Background subtracted light curves averaged over 10 seconds are shown for Cyg X-1 observation for LAXPC detector A in Panel (a) and for LAXPC detector B in Panel (b) for the period 4.06-7.25 am on 2008 April 14.}
\label{fig12}
\end{figure}
It can be seen from Figure \ref{fig12} that an excess count rate of $\sim$ 150 counts per second above the average background was detected from Cyg X-1 in each detector. Intensity variations of Cyg X-1 are clearly observed in the hard X-rays ($>$ 20 keV) in both the detectors. Rapid X-ray variability over time scale of tens of seconds and minutes is apparent in the light curves of both the detectors. It should be pointed out that very large variations more than a factor of $\sim$ 5, be treated with a caution as they may be due to offset in source pointing. As indicated earlier some times gondola was oscillating with $\sim$ 30 sec period and though we had identified and removed these data, we may not have done it fully which will give rise to large variability. We attribute nearly zero count rates in the light curve to the large pointing offset from the source position. The time scale and magnitude of variations in both the detectors are in broad agreement. Source count rate of about 150 counts per sec implies that one can construct meaningful Power Density Spectrum (PDS) of bright X-ray sources like Cyg X-1 to study Quasi-Periodic Oscillations (QPO's) in hard X-rays which is one of the objectives of the LAXPC instrument on the ASTROSAT mission. The LAXPC instrument on ASTROSAT has 3 LAXPC detectors with total effective area of $\ge$ 6000 cm$^{2}$ in 15-60 keV range and equipped with 1$^{\circ}$$\times$1$^{\circ}$ field of view collimators \cite{Agrawal 2006a}.
 
At satellite altitude the source count rate will be much higher as there is no absorption of 20 keV photons due to atmosphere and the background count rate is also expected to be lower. Most X-ray sources emit a dominant part of their flux below 20 keV due to steep energy spectra. At satellite altitude most of the counts will arise due to X-rays of $<$ 20 keV, as a result studies of QPO's and other types of rapid intensity variations will be possible with high sensitivity even for sources at $\sim$ 1 milliCrab. A PDS was generated using the integral count rate data obtained from detector A. We have subtracted the Poissonian noise level from the PDS. Data from 11279 sec observation of Cyg X-1 during 4.06-7.25 am were used to construct the PDS using XRONOS software (Heasarc). The PDS of Cyg X-1 in the frequency range of 0.01-0.5 Hz with a bin time of 1 second is shown in Figure \ref{fig13}. 

\begin{figure}
\centerline{
\includegraphics*[width=6.0cm, angle=270]{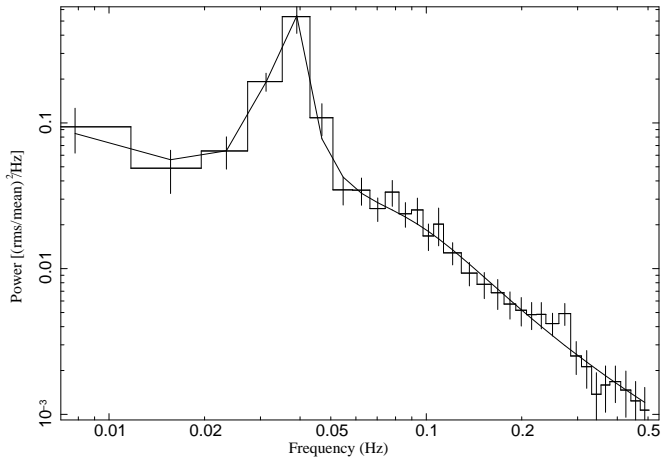}}
\caption{Power density spectrum of Cyg X-1 obtained using 4.06-7.25 am data from LAXPC detector A. Power is in unit of (rms/means)$^{2}$/Hz.}
\label{fig13}
\end{figure}

 The shape of the observed PDS is consistent with the low/hard state of Cyg X-1 (\cite{Paul 1998}). The PDS were fitted with a power-law model, standard models from XRONOS software. The PDS has power law shape with power-law photon index of -1.06. No QPO feature is detectable in the PDS of Cyg X-1. The power was found to increase towards the lower frequency from 0.1 Hz. At $\sim$ 0.03 Hz a peak is detected in the PDS due to the periodicity of oscillations of the gondola during the pointed of Cyg X-1 observation. The PDS in Figure \ref{fig13} steepens above 0.05 Hz.

\subsection{X-ray Spectrum of Cygnus X-1}
To derive the energy spectrum of Cyg X-1 we generated response matrix for the LAXPC detector A using a Monte Carlo routine. Detector characteristics such as Xenon gas pressure, absorption due to window material, energy resolution of the detector, geometry of the detector and the characteristics of the event selection logic such as lower and upper level energy thresholds, production probability of Xe-K-fluorescent photon etc serves as input to the program. This program assumes random incidence of photon at the top of the collimator. Simulation was performed 10000 times for each energy bin from 2-115 keV using all anodes and all layers together. For the response matrix the upper limit for simulation is taken as 115 keV. A photon of 115 keV will appear as a 80 keV photon due to escape of 29.8 keV Xenon fluorescent X-rays. We used data from second stretch of Cyg X-1 observation (6.26-7.25 am) for generating the count rate energy spectrum  of Cyg X-1 in 3-80 keV energy range.   

The incident spectrum of Cyg X-1 impinging on the top of the detector was  produced by absorbing it through 2.5 g/cm$^{2}$ of residual atmosphere and then folding it through detector response function. The spectral data is rebinned in 32 channels. 
A single temperature thermal Compton model is fitted to the observed count rate spectrum of Cyg X-1. The thermal Comptonization model involves inverse Compton scattering of low energy photons by hot thermal electrons in the corona of an accretion disk around a black hole. 
The Tuebingen-Boulder ISM absorption model (TBabs model from XSPEC software\footnote{https://heasarc.gsfc.nasa.gov/xanadu/xspec/manual/XSmodelTbabs.html} and a single temperature Comptonization  model (compST model from XSPEC software\footnote{https://heasarc.gsfc.nasa.gov/xanadu/xspec/manual/XSmodelCompst.html}), are used to fit the spectrum. CompST model was used for spectral fitting of Cyg X-1 data obtained from a balloon borne experiment data by Chitnis et. al. 1998 (\cite{Chitnis 1998}). The value of nH=6$\times 10^{21}$ atoms cm$^{-2}$ in TBabs is obtained from Frontera et al. 2001 \cite{Frontera 2001}.   

The best fit spectrum of Cyg X-1 is shown by a thick black line in Figure \ref{fig14} along with the observed count rate giving a reduced $\chi^{2}$ of 1.1. A 5\% systematics is considered in the spectral fitting to account for the instrumental effects. The spectral parameters of the best fit spectrum are temperature, kT$_{e}$=25.4$^{+2.1}_{-1.7}$, optical depth ($\tau$)=15.3$^{+1.1}_{-1.0}$ with normalization constant, norm=3.4$^{+0.4}_{-0.4}$$\times$10$^{-5}$ counts cm$^{-2}$s$^{-1}$.

It may be pointed out that Cyg X-1 was in a low-hard state on 2008 April 14 as seen from the light curve of All Sky Monitor (ASM\footnote{http://xte.mit.edu/XTE/ASM\_lc.html}). The derived spectrum is consistent with the hard state of Cyg X-1.  
%}

\begin{figure}
\centerline{
\includegraphics*[width=8.0cm, angle=270]{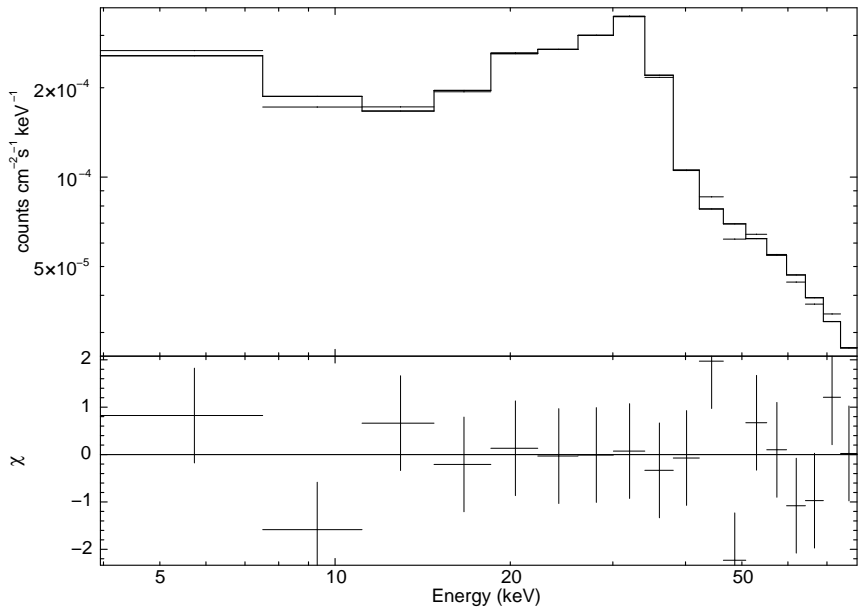}}
\caption{The energy spectrum of Cyg X-1 observed count rate along with the folded model, in 3-80 keV obtained using 6.26-7.25 am data from LAXPC detector A.}
\label{fig14}
\end{figure}

\section{Conclusion}
The main objective of LAXPC Balloon Flight Experiment was to validate the design of the LAXPC detectors by testing their performance in space-like environment. An additional aim was to measure the detector background rates and assess sensitivity for studying cosmic sources. The experiment also provided assessment of performance of the HV units, signal processing electronics of detectors and data modes of LAXPC instrument. The LAXPC detectors and associated electronic systems performed as expected during the balloon flight.  The measured background rates are consistent with those estimated from an independent Monte Carlo simulation of the detector response (unpublished). A timing accuracy of 10 $\mu$s in the time tagging mode of data acquisition during the balloon-borne experiment was realized.  This is necessary to study kHz QPO and other types of rapid variability. Results from the balloon experiment validate the design of the LAXPC instrument and it is hoped that this instrument will fulfil its science objectives.

\begin{acknowledgements}
We are extremely grateful to the engineering and technical staff of TIFR Workshop who carried out fabrication of major parts of LAXPCs and gondola system of the instrument. We express our sincere thanks to the TIFR Balloon Facility staff under the supervision of Mr S. Sreenivasan who carried out all the balloon flight related operations in an excellent manner. It is a pleasure to acknowledge the help of Prof H. M. Antia of TIFR in generating detector response matrix. We are grateful to late Mr. M.R.Shah and Mr. K. Kutty  for their contributions to the orientation system and its fabrication used in this experiment. This research was in part supported by NASI that awarded Senior Scientist Fellowship to P. C. Agrawal. We also thank the anonymous referee for their suggestions which improved the manuscript.
\end{acknowledgements}

\end{document}